\documentclass[useAMS,usenatbib,usegraphicx,aastex]{mn2e}

\title{Early polarization observations of the optical emission of gamma-ray bursts: GRB\,150301B and GRB\,150413A.}

\author[E. S. Gorbovskoy et al]{E. S. Gorbovskoy$^{1,2}$
\thanks{E-mail:gorbovskoy@sai.msu.ru,  lipunov2007@gmail.com},
V. M. Lipunov$^{1,2}$,
D.A.H. Buckley$^{3}$, 
V. G. Kornilov$^{1,2}$, \newauthor
P. V. Balanutsa$^{1,2}$,                           
N. V. Tyurina$^{1,2}$,
A. S, Kuznetsov$^{1,2}$,
D. A. Kuvshinov$^{1,2}$,  \newauthor
I. A. Gorbunov$^{1,2}$,
D. Vlasenko$^{1}$,
E. Popova$^{1,2}$,
V. V. Chazov$^{1,2}$, 
S. Potter$^{3}$,
M. Kotze$^{3}$,\newauthor
A. Y. Kniazev$^{3,11}$,
O. A. Gress$^{4}$,
N. M. Budnev$^{4}$, 
K. I. Ivanov$^{4}$, 
S. A. Yazev$^{4}$, \newauthor
A. G. Tlatov$^{5}$, 
V. A. Senik$^{1,2,5}$
D. V. Dormidontov$^{1,2,5}$, 
A. V. Parhomenko$^{1,2,5}$, \newauthor
V. V. Krushinski$^{6}$, 
I. S. Zalozhnich$^{6}$, 	
R. Alberto Castro-Tirado$^{7}$,
R. S\'anchez-Ram\'irez$^{7}$, \newauthor 
Yu. P. Sergienko$^{8}$, 
A. Gabovich$^{8}$, 
V. V. Yurkov$^{8}$,
H. Levato$^{9}$,  \newauthor
C. Saffe$^{9}$,
C. Mallamaci$^{10}$, 
C. Lopez$^{10}$,
F. Podest$^{10}$\\
$^{1}$Lomonosov Moscow State University,GSP-1, Leninskie Gory, Moscow, 119991, Russia\\
$^{2}$Sternberg Astronomical Institute, Moscow State University,  Universitetskiy pr. 13, Moscow 119992, Russia\\
$^{3}$South African Astronomical Observatory, PO Box 9, 7935 Observatory , Cape Town, South Africa \\
$^{4}$Irkutsk State University, ul. Karla Marxa 1, Irkutsk 664003, Russia\\
$^{5}$Kislovodsk Solar Station of the Pulkovo Observatory RAS, P.O.Box 45, ul. Gagarina 100, Kislovodsk 357700, Russia\\
$^{6}$Kourovka Astronomical Observatory, Physical Department of Ural State University, pr. Lenina 51, Ekaterinburg  620083,  Russia\\
$^{7}$Instituto de Astrof\'isica de Andaluc\'ia (IAA-CSIC), Glorieta de la Astronom\'ia s/n, 18008 Granada, Spain \\
$^{8}$Blagoveschensk Educational State University, ul. Lenina 104,  Amur Region, Blagoveschensk 675000, Russia\\
$^{9}$Instituto de Ciencias Astronomicas, de la Tierra y del Espacio, Av.España Sur 1512, J5402DSP, San Juan, Argentina\\
$^{10}$Observatorio Astronómico Félix Aguilar, San Juan - Argentina.\\
$^{11}$Southern African Large Telescope Foundation, PO Box 9, 7935 Observatory, Cape Town, South Africa
}

\begin{document}


\pagerange{\pageref{firstpage}--\pageref{lastpage}} \pubyear{2000}

\maketitle

\label{firstpage}

\begin{abstract}

We report early optical linear polarization observations of two gamma-ray bursts made with the MASTER robotic telescope network. We found the minimum polarization for GRB\,150301B to be 8\% at the beginning of the initial  stage, whereas we detected no polarization for GRB\,150413A either at the rising branch or after the burst reached the power-law afterglow stage.
This is the earliest measurement of the polarization (in cosmological rest frame) of gamma-ray bursts.
The primary intent of the paper is to discover optical emission and publish
extremely rare (unique) high-quality light curves of the prompt optical emission of gamma-ray 
bursts during the non-monotonic stage of their evolution. We report that our team has discovered the optical counterpart of one of the bursts, 
GRB\,150413A.

\end{abstract}


\section{Introduction}
\label{section:Introduction}

Gamma-ray bursts are among the most distant and powerful explosions in the Universe. 
	The idea that gamma-ray burst emission originates in a narrow magnetized jet with an opening angle of ~ 1 degree now appears to be a generally accepted hypothesis, though so far unsupported by direct experimental evidence. One of the consequences of the presence of jets and ordered magnetic fields in these jets could be observation of the polarization of the prompt emission of the burst. Prompt optical emission of gamma-ray bursts is the emission that originates at the time when the central engine of the burst still continues to operate, i.e., when the gamma-ray emission has not yet faded. 
 
Scarce and always unique (one burst, one telescope) observations of early optical emission, which is sometimes synchronous with gamma-ray emission, show that bursts may exhibit one of the two behavior patterns \citep{Vestrand2005}: 
\begin{enumerate}

\item optical emission appears simultaneous with gamma-ray emission (or sometimes even before it)and appears correlated with the gamma-ray lightcurve and \item optical emission appears at the end of the gamma-ray pulse and shows no correlation in form with the hard radiation. This dichotomy has been confirmed by later observations of the gamma-ray bursts GRB\,100901A and GRB\,100906A \citep{Gorbovskoy2012}. 
\end{enumerate}
	Several authors have reported observing 10 to 25 percent polarization of gamma-ray emission \citep{Granot2003,Yonetoku2011}. In addition, numerous attempts were made in the late 20-th and early 21-th century to detect the polarization of the optical afterglow 10 --- 20 hours after the burst. It turned out that late afterglow either shows no polarization or only a weak polarization at a level of several percent \citep{Covino2004}.
	The authors of several early polarization observations of gamma-ray bursts \citep{Steele2009,Uehara2012,Kopac2015,Mundell2007,Mundell2013,King2014,Cucchiara2011} reported detecting polarization at a 10\,\%  level. Also GRB afterglows can be circularly polarized at optical wavelengths \citep{Wiersema2014}.
	However, so far not a single positive detection has been made of the polarization of the prompt optical emission of the gamma-ray burst. The problem is with the duration of gamma-ray bursts, which are too short and usually last no more than several tens of seconds. Hence fully automatic observations are needed to be made using robotic telescopes equipped with polarimetry devices. 	 
	To perform early polarimetric observations of the prompt optical emission of gamma-ray bursts, we developed \citep{lipunov_etal2010,Lipunov2004AN}  a network of MASTER II robotic telescopes\footnote{http://observ.pereplet.ru/}, which we installed in seven sites across the globe \citep{Kornilov2012}. Each telescope has two wide-field (4 square degrees each) 400-mm diameter optical tubes\footnote{It looks like a binocular, one telescope to the east and one telescope to the west sides of the mount.} and is equipped with an extra axis on a superfast mounting capable of moving at a speed of 15-30 degrees/s and aligning the tubes in parallel when receiving an alert from cosmic gamma-ray observatories. Each camera is equipped with a linear polarizer with mutually perpendicular polarization planes. The axes at the neighboring telescopes of MASTER network are turned by 45 degrees relative to each other. Therefore for complete determination of the polarization the telescopes at two neighboring sites have to be pointed to the same target simultaneously. However, recording the flux difference between the two perpendicular polarization directions yields a lower limit for the polarization degree of the emission \citep{Pruzhinskaya2014}. The pointing, exposure, and image processing are performed in automatic mode \citep{Gorbovskoy2013}.
In the time when MASTER is not observing gamma-ray bursts it performs a sky survey, which has resulted in the discovery of more than 800 optical transients of nine astronomical types in the last three years.

In this paper we report the results of early polarimetric observations of the optical emission (including the prompt optical emission that varies in concert with gamma-ray emission) at the early non monotonic light curve stage, which then transforms into standard self-similar power-law decrease.

\section {Observation technique}
\label{section:obs}

\subsection{Polarimeter}
\label{section:pol}

The MASTER net was designed with the objective to deliver polarization information as early as possible after GRB triggers. More than 100 observations of GRBs were made by the MASTER global robotic net. Optical emission was detected for $\sim$ 20 GRBs \citep{Lipunov2007ARep,Gorbovskoy2012,Pruzhinskaya2014}. The GRB\,100906A, GRB\,110422A, and GRB\,121011A events deserved attention because their optical observations were carried out during the gamma-ray emission. 

We have a linear polarimeters only. So at this article we assume zero circular polarization.

Let $I_1$ and $I_2$ be the  signals of the detector with orthogonal polarimeter filters (for example 
$45^o$ and $135^o$ interposed). Then a value $P_{lowlim}=(I_1-I_2)/(I_1+I_2)$ be a lower limit of linear polarization. 

To derive value of polarization level  one needs to perform observations with filters
for linear polarization positioned at three angles, for example, at
$0^o$, $45^o$, and $90^o$ with respect to the reference direction. We
have  two orthogonal filters at each of the MASTER-II telescopes; thus, if
such telescope observes alone, only a lower limit on polarization level --- $P_{lowlim}$ can be
deduced.


If the module of $P_{lowlim}$ is less than measurement error $\sigma_\mathrm{p}$ , then formally we
have zero as a minimum estimate for the degree of linear polarization.
100-percent linear polarization is also possible in this case, if a
filters is accidentally positioned so that a source's polarization
plane is inclined by $45^o$ to 
both of them.

The two synchronous frames (taken with different cameras) used to measure $P_{lowlim}$ are mutually calibrated so that the average $P_{lowlim}$ for comparison stars would be $ Avg (P_{lowlim}) = 0$. 
This is achieved at the stage of the photometric calibration. We use the same reference stars for frames in both polarizations. The USNO B-1 catalog that we use is naturally produced in unpolarized light. 
It means that $ Avg (I_{45} - I_{135}) = 0 $ and, consequently $ Avg (P_{lowlim})=0 $. 
Of course with such a calibration we zero average polarization of the background stars. I.e. Galactic polarization.  However, in the context of this work it is convenient even  the object of study is the polarization of extragalactic sources and Galactic polarization  must be excluded in any case. Of course this can only work under the assumption of uniform and equal polarization of the background stars. Given the large distance from the galactic plane ( galactic latitude ($l$) for bursts is: $l_{\rm GRB150301B} = -29.8^d$ and $l_{\rm GRB150413A} = 45.5^d$ ), not huge Galactic absorption ( $A_V = 0.20$ for GRB\,150301B and  $A_V = 0.05$ for GRB\,150413A \citep{Schlafly2011} ) and small area covered by reference stars ($0.13 \times 0.13$  deg.) 
we believe it is possible to work in such an approximation.

In the case of the two GRBs considered below MASTER optical linear polarization observations were made only in two polarizers with no significant difference found in the two channels for any of the events. Unfortunately, it is impossible to make a conclusion about the absence of polarization if observations only in two polarizers are available (the polarization plane may be aligned at 45 degrees relative to the polarizers).

All MASTER magnitudes reported at this paper is unfiltred magnitude calibrated as $ 0.2B + 0.8R $ by the USNO-B1 catalog.  The reasons for using this calibration are discussed in the early papers \citep{Kornilov2012,Gorbovskoy2012}. MASTER polarization band is determined by the respond curves of the CCD camera and a transmission curve of polarizing filter, which have been reviewed here \citep{Pruzhinskaya2014}.

\subsection{Determination of an errors}
\label{error_det}
In the observation of gamma-ray bursts two factors need to be considered:
$a)$ we needed a high temporal resolution  in the initial stages of the light curve and $b)$ it is important not to loose a quickly decaying source from view.
So MASTER exposure times follow the relation $t_{exp} = (T_{start} - T_{trigger})/5$, where $T_{trigger}$ is the trigger time (UT), $T_{start}$ is the time of the beginning of exposure (UT). The exposure time is rounded to an integer with a step of 10 s and cannot exceed 3 min. In consequence, frames are quite different from each other and the task of  error determination becomes difficult.  To do this, we analyze the distribution of $P_{lowlim}$ of the comparison stars, depending on the magnitude. 
Then, we select only the stars with magnitude  in the interval $m_{grb} \pm 0.5 $, where $m_{grb}$ is a GRB magnitude on a given frame.  Dispersion of comparison stars $P_{lowlim}$  in this interval we call as $1 - \sigma $ error of GRB $P_{lowlim}$.
We use the comparison star in a radius of $0.25^o$ for GRB\,150301B and $1^o$ for GRB\,150413A.  
This difference is due to the bit high value of the absorption in the case of GRB\,150301A and a desire to meet the conditions specified in section \ref{section:pol}.
Any errors that are reported in this paper is  $1 \sigma$.

\section{Measurement of early polarization of the gamma-ray burst GRB 150301B.}
\label{s:150301B}

The gamma-ray burst GRB\,150301B was recorded and localized by Swift space observatory at 19:38:04 UT on March 1, 2015 \citep{gcn17515}. According to the data from Swift/BAT instrument, GRB\,150301B was a long gamma-ray burst with $T_{90} = 12.44 \pm 1.49 sec$  \citep{gcn17524}.  



MASTER-SAAO robotic telescope received the socket message from the GCN network, immediately interrupted its regular survey, pointed to the burst localization area, and started observations in two polarization filters 60 sec after notice time and 79 sec after trigger time at 2015-03-01 19:39:23 UT. The very first frame showed a bright ($15.3^m$) optical source with the coordinates 
\begin{eqnarray} 
RA(J2000) &=&\quad05h\quad 56m\quad 39.94s \nonumber \\ 
DEC (J2000) &=&\quad-57d\quad 58m\quad  10.0s \nonumber 
\end{eqnarray} 

and an error of 1 arc. second \citep{gcn17518}. 

The Swift XRT X-ray telescope and  UVOT ultraviolet telescope also detected X-ray and optical emission, respectively, from  GRB\,150301B 82 and 85 seconds after the BAT trigger  \citep{gcn17515}. 
Later, $5.12^h$ after the burst, the Very Large Large Telescope (Paranal Observatory, Chile) and X-shooter measured the redshift of this burst, which was found to be  $z=1.5169$ \citep{gcn17523} . 
MASTER observations. 
MASTER robotic telescope observed this gamma-ray burst in the alert mode in two polarization planes with variable exposures (see section \ref{error_det}). 

\begin{figure}
\includegraphics[width=84mm]{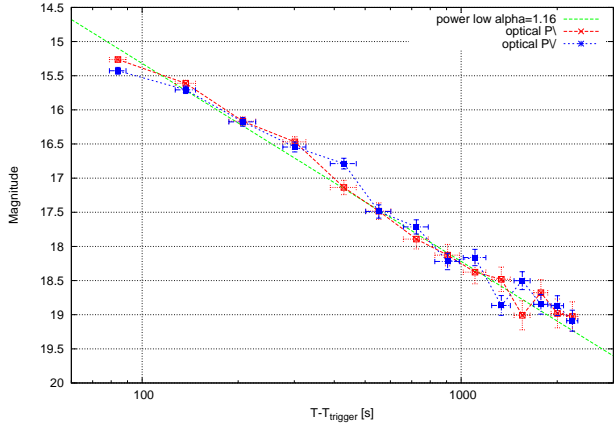}
\caption{
GRB\,150301B MASTER-SAAO  light curve in two telescope tubes with different polaroid orientations (45 and 135 deg).
\label{fig:grb150301B_lc}}
\end{figure}

As a result of our observations the following light curve was obtained (see Fig.\,\ref{fig:grb150301B_lc} and Table\,\ref{tab:polarization150301B} ).

MASTER-SAAO telescope observations of GRB\,150301B lasted from 2015-03-01 19:39:23 UT through 2015-03-01 20:48:11, i.e., for about 1 hour beginning with the 79th second after the trigger. Observations were then interrupted because of heavy clouds. 
We observed well-defined optical emission from GRB 150301B up until  ~ $2000$ sec after the trigger while the source remained brighter than the limiting magnitude on our frames. Unfavorable weather conditions prevented us from extending the light curve by obtaining deep stacked frames. Throughout the entire time interval from  $79$ to $2000$ sec the source faded in accordance with the power law $ F \sim t^{-1.16 \pm 0.05} $ which, in general, is typical of GRB that time \citep{meszaros2006}.

\subsection{Spectrum from the optical to X-ray.}

\begin{figure}
\includegraphics[width=84mm]{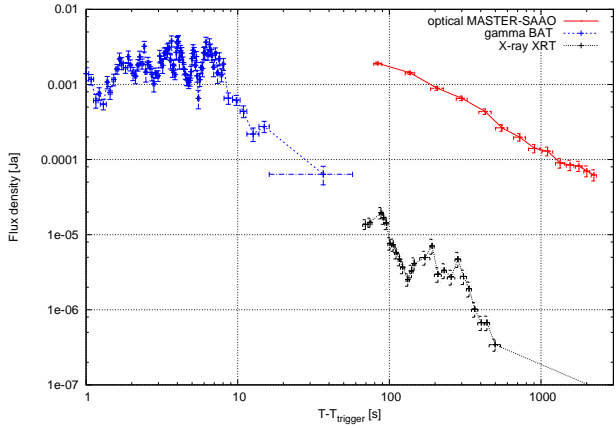}
\caption{
Optical, X-ray and gamma observation light curve of GRB\,150301B.\label{fig:grb150301B_spectra}
The {\em Swift}/XRT 0.3-10~keV and {\em Swift}/BAT 15-150~keV light curves are available 
from the {\em Swift} Burst Analyser
~\citep{evans_etal2010}
(\textit{http://www.swift.ac.uk/burst\_analyser/}).
}
\end{figure}

Observations of the gamma-ray burst with MASTER telescopes were performed synchronously with observations on the  Swift/XRT X-ray telescope (Fig.\,\ref{fig:grb150301B_spectra}). 
Optical data corrected for absorption   $N_H= 6.2 \times 10^{20} cm^{-2}$ \citep{gcn17533} which is  consistent with the Galactic value \citep{Willingale2013}.
Note that variability observed in X-rays in the form of three local maxima in the $88$, $191$ and $282$ seconds does not occur in the optical wavelength, that has already been observed previously (for example \citep{Ziaeepour2008,Li2012}).

\subsection{Polarimetric measurements. }\label{section:polarimertic_measure}

The technique of polarimetric measurements used on MASTER network telescopes allows detecting the optical emission polarization at a level of  5\% or higher for objects brighter than 16m or estimating a lower limit for polarization (if the object is observed only by one of the observatories of the network) \citep {Pruzhinskaya2014}.

As is evident from the light curve (Fig\,\ref{fig:grb150301B_lc}), the data points for different polarizers differ substantially for the first and fifth observations suggesting possible polarization of the emission. 

\begin{figure}
\includegraphics[width=84mm]{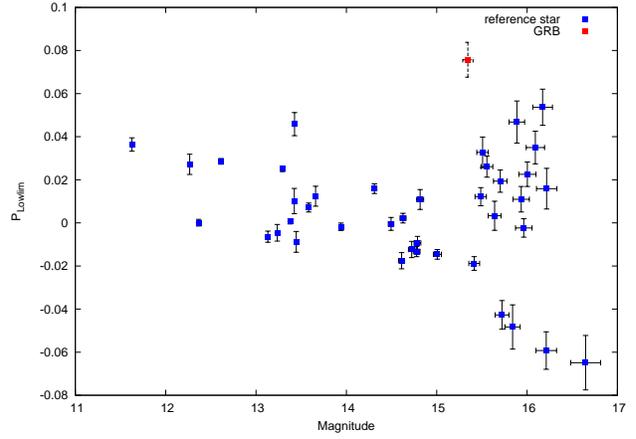}
\caption{
This diagram shows the distribution of $P_{lowlim}  $ depending on the magnitude of  GRB\,150301B and reference stars at a first  MASTER image.
 This figure illustrates the method for $ P_{lowlim} $ error determining, described in section \ref{error_det}. 
As $1 \sigma$ error of the object we call the polarization dispersion of reference stars in the range of $ \pm 0.5$ magnitude relative to the magnitude of the object.
In this case the object has a $ P_{lowlim} = 7.6 \pm 0.8 $  with  $3.2 \sigma$  significance level.
Here and in a similar figures below  $ P_{lowlim} $  formaly could by a negative. 
 As discussed above $ P_{lowlim} = (I_{45} - I_{135}) / (I_{45} + I_{135}) $. If the $ I_{135} $> $ I_{45} $, we obtain a formal negative $ P_{lowlim} $. Physically, this comes from the fact that we can not determine the angle of polarization. However, the positive value of $ P_{lowlim} $ indicates that $ 0 <PA < 90$, and  a negative that $ 90 <PA <180 $, where $PA$ is a polarization angle. 
\label{fig:grb150301B_error_1frame}}
\end{figure}

\begin{figure}
\includegraphics[width=84mm]{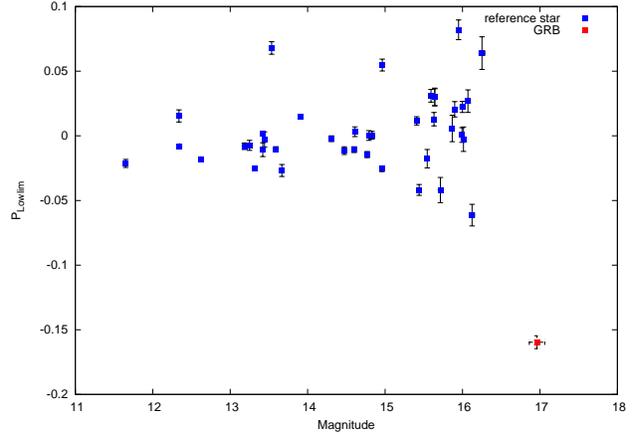}
\caption{
$ P_{lowlim} $ vs. stellar magnitude for GRB\,150301B and reference stars on fifth image. Although technically significant value of $ P_{lowlim} $ there are no reference start with magnitude lower than  GRB\,150301B magnitude on this frame. Region $ m_{GRB150301B} \pm 0.5 $ is not fully covered by the reference stars. Therefore, we can not determine the statistical significance of the measure and, accordingly, be considered that this value is real. 
\label{fig:grb150301B_error_5frame}} 
\end{figure}

In this case observations were performed only with MASTER-SAAO telescope and we now have at our disposal synchronous observations in two polarization planes at 45 and 135 degrees. As can be seen from the light curve in the 1st and 5th of exposure (79 and 389, respectively), there is a significant difference between the measurements at different polarizations, which suggests a possible polarization of radiation at a given time.
Therefore for  GRB\,150301B we can estimate a lower limit for the polarization degree of its optical emission, which we compute by the following simple formula: 
$ P_{lowlim} = (I_{45} - I_{135}) / (I_{45} + I_{135}) $ ,
where $P_{lowlim}$ is the lower limit of polarization, i.e., $ P_L >= P_{lowlim}$,  and $I_{45}$  an $I_{135}$ are the optical fluxes recorded through a linear polaroid turned by $45$ and $135$ degrees, respectively. 
Figure\,\ref{fig:grb150301B_error_1frame} shows the magnitude dependence of $P_{lowlim}$ for GRB\,150301B and comparison stars. All reference stars were selected within the 256 pixel or 8 arcmin radius from the gamma-ray burst. Stars with cosmetic defects on one of the CCDs  (broken and hot pixels) were removed manually. The two synchronous frames (taken with different cameras) used to measure $P_{lowlim}$ are mutually calibrated so that the average $P_{lowlim}$ for comparison stars would be $ Avg (P_{lowlim}) = 0$. 
To determine the error of the inferred lower polarization limit for GRB 150301, we measured the standard deviation of $P_{lowlim}$ from zero for the comparison stars in the magnitude interval  $M_{GRB150301B} \pm 0.5$, i.e., from $14.9$ to $15.9$.
The lower limit for the polarization of  GRB\,150301 during the time interval from $79$ to $89$ s is thus equal to $P_{GRB150301B} > 7.57 \pm 0.04 $ at a $3.2 \sigma$ significance level. 
Despite the fact that formally the polarization of the 5-th point (~400 sec after the burst) is great, but the object is visible almost at the limit of detection as shown in Fig.\,\ref{fig:grb150301B_error_5frame} can not find a suitable comparison star magnitudes.

\section{Observations of GRB 150413A.}
\label{obs150413A}

This gamma-ray burst was detected by BAT instrument of Swift observatory \citep{gcn17688}.
At 13:54:58 UT, the Swift Burst Alert Telescope (BAT) triggered and located GRB\,150413A . 
Unfortunately due to an observing constraint X-ray observations of GRB\,150413A were not made. 

The telescope of MASTER robotic network located at Tunka astrophysical center near Baikal lake was automatically pointed to the target 50 seconds after receiving the alert message and 132 seconds after the trigger operated onboard SWIFT space observatory.  

\begin{figure}
\includegraphics[width=84mm]{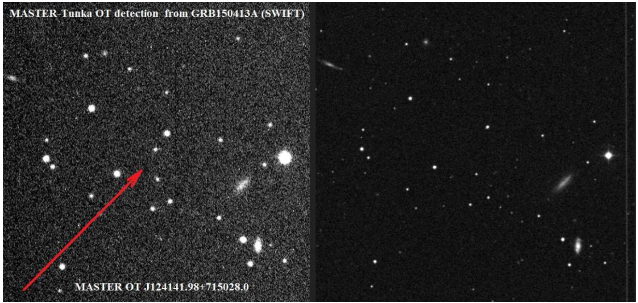}
\caption{
Optical burst within the error  box of the gamma-ray burst GRB\,150413A.\label{fig:grb150413A_img}}
\end{figure}

We found nothing new in the first frame within the error box of SWIFT BAT telescope. However, on  the second and next frames a hitherto unknown source was found by MASTER optical transient autodetection system at a position 
\begin{eqnarray}
RA(J2000) &=& \quad 12h \quad 41m  \quad 41.98s \nonumber \\ 
Dec(J2000) &=& \quad +71d \quad 50'\quad  28.0'' \nonumber 
\end{eqnarray}
and with uncertainty 1 arcsec and unfiltered magnitude about $m = 16.1^m$  \citep{gcn17689} . (Fig.\,\ref{fig:grb150413A_img})

Its brightness increased steadily and by the $400-th$ second it the object became brighter than $15^m$ \cite{gcn17690}. We seemed to have found an unusual gamma-ray burst, which remained brighter for longer than in most previous events. 


However, following the MASTER detection, the 2.2m telescope at the German-Spanish Calar Alto (CAHA) Observatory observed the afterglow starting at 20:25 UT. (i.e. 6.5 hr post-burst) under poor weather and seeing conditions \citep{gcn17697}. Data was reduced and calibrated on the usual way using IRAF and custom tools coded up in python. The resulting spectrum has a low SNR ratio, but we clearly detect a broad absorption line which we identify as Lyman-alpha. No metal line is significantly detected in the wavelength range covered due to the observing conditions. We performed a Voigt profile fitting using Ly$\alpha$ and Ly$\beta$ lines, finding a value of  $ z = 3.140 \pm 0.005 $ for the redshift and $ log N_H = 22.0 +- 0.25 $  for the HI column density. This redshift is well constrained due to the identification of the Ly$\beta$ line, and consistent with the later provided by de Ugarte Postigo and Tomasella \citep{gcn17710}. The value of the HI column density we find is hardly seen in QSO sight-lines, but quite common in GRB-DLAs (e.g., \citep{Cucchiara2015}).

So everything became clear.  If the  gamma-ray burst had a redshift $ \sim 3.1$ \citep{gcn17697,gcn17710}, it means that all processes appear slowed down by a factor of  $(1+z) ~ 4$. With this correction applied, we converted our observations to the comoving reference frame of the object and found the event to have a proper duration of less than 1 minute and hence to be a rather typical long gamma-ray burst most likely resulting from the formation of a black hole after the core collapse of a massive star. 
	
\begin{figure}
\includegraphics[width=84mm]{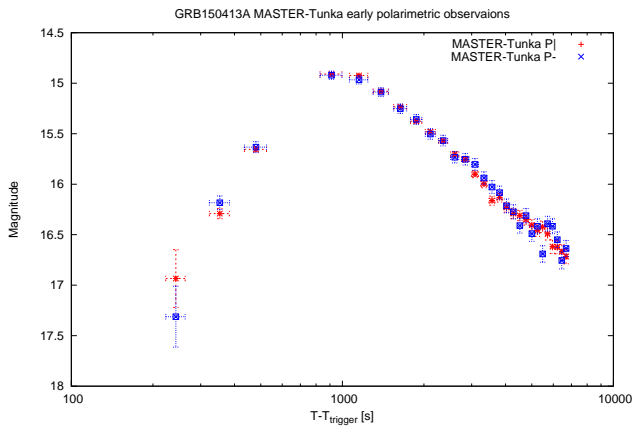}
\caption{
GRB\,150413A MASTER-Tunka  light curve in two telescope tubes with different polaroid orientations (0 and 90 deg).\label{fig:grb150413A_lc}}
\end{figure}

The light curve of GRB\,150413A shown in Fig.\,\ref{fig:grb150413A_lc}. This gamma-ray burst, as the previous one was observed only by  one telescope of MASTER  net --- MASTER-Tunka. The observations at the telescope MASTER-Tunka were made in two polarization angles (0 and 90 degrees, respectively). Therefore, as in the previous case, we can talk about the lower limit of the possible polarization of the afterglow of GRB\,150413A.

\begin{figure}
\includegraphics[width=84mm]{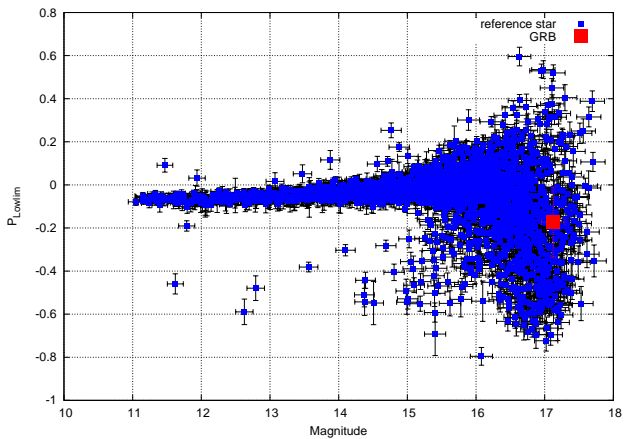}
\caption{
Value of the polarization low limit ($P_{lowlim}$) as a function of the magnitude on the second frame of GRB\,150413A observations. 
Explanations asymmetry present image and fig.\,\ref{fig:grb150413A_err15} given in chapter \ref{obs150413A}.
\label{fig:grb150413A_err1}}
\end{figure}

Figure\,\ref{fig:grb150413A_err1} is a graph of the degree of polarization of GRB and background stars in the first frame. Diagram plot on the assumption that all the brightest stars (brighter than magnitude 15) in the frame have zero polarization. Following this assumption,  the median value of $ P_{lowlim} = (I_1-I_2) / (I_1 + I_2) $ sample of stars brighter than 15  magnitude were normalized to zero. In general (for more than 95\% of the stars), the diagram keeps symmetry up to the  limit at about $ 17^m $, with increasing spread is inversely proportional to the magnitude of star. This spread will characterize the measurement error.  Huge scatter for the bright stars (on the level of polarization like 40\%) related to the false identification of closely spaced stars. Strong variations in  $ m > 17^m $  region connected with small difference of  the upper  limits of two frames in different polarizations.

\begin{figure}
\includegraphics[width=84mm]{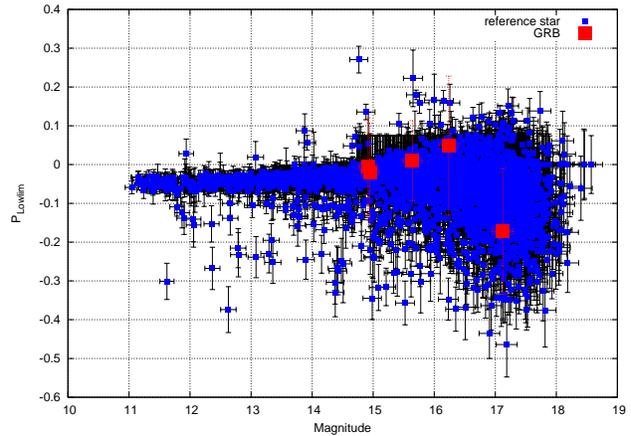}
\caption{
Value of the polarization low limit ($P_{lowlim}$) as a function of the magnitude on the first five frames. All GRB points covered by reference stars. So there is no evidence to believe that there is a polarization for this GRB \label{fig:grb150413A_err15}}
\end{figure}

As can be seen from Figure\,\ref{fig:grb150413A_err1}, despite formally a significant degree of polarization of the first measured point ($ P_{lowlim} ~ 15\% $), it is less than the measurement error. That is not surprising since the burst in this picture has a very close to limiting magnitude. On all subsequent (till a maximum  burst luminosity) frames polarization also not found, because all polarisation measurements fall within the basic distribution of background stars, as seen on Fig.\,\ref{fig:grb150413A_err15}. It shows all the background stars with an average polarization for the first five images and points represent gamma-ray burst on these frames. Diagram is plotted on the same principle as the previous one.

\section{Discussion.}

As we already pointed out, \citep{Vestrand2005} noted that two behavior types of early optical emission are observed for gamma-ray bursts. In some cases optical emission varies in concert with gamma-ray flux, whereas in other cases the optical light curve is totally uncorrelated with the gamma-ray flux and usually consists of a smooth rise followed by power-law afterglow. We earlier observed both types of such behavior \citep{Gorbovskoy2012} . 
It is logical to suggest \citep{Vestrand2005} that in the former case we are dealing with the optical emission arising in the internal shock region, where it is generated simultaneously with the gamma-ray emission. In this case the optical light curve reflects the internal properties of the relativistic magnetized jet (ultra-relativistic ejecta driven by the GRB ) and detection of its polarization would provide unique information about the jet structure.
The optical emission of the second type may be associated with a different region of the jet (in the standard fireball model \citep{meszaros_rees1999}), which forms as a result of relativistic jet colliding with interstellar gas compressed in the bow shock or with the stellar wind of the progenitor.

\begin{figure}
\includegraphics[width=84mm]{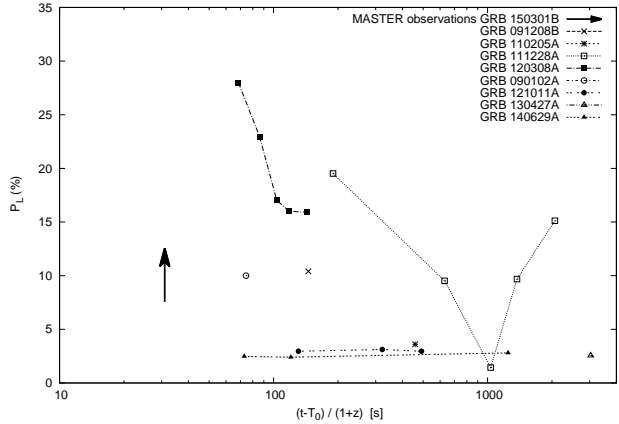}
\caption{
Following figure 3 at \citep{Mundell2013}   we plot MASTER polarization low limit in comparison with other known earlier measurements of the GRB polarization. The figure shows the degree of optical
polarization $P_L$  as a function of time
after the burst in the cosmological rest frame,
$(t-T0)/(1+z)$, where $z$ is redshift. Redshift for GRB\,150301B is $ z = 1.5 $ and measured by \citep{gcn17523}. 
The data for GRB\,111228A is preliminary (from a conference talk K. Takaki \citep{gcn12787}
(\textit{http://nagataki-lab.riken.jp/workshop/SNGRB2014/takaki.pdf}).
).
\label{fig:all_polarizations}
}
\end{figure}

The gamma-ray bursts that we report in this paper belong to the second type. It is for such bursts that earlier polarization was observed at the afterglow stage. 
Our observations very much confuse the problem of the optical emission polarization of gamma-ray bursts. The point is that we detected polarization only for the GRB\,150301B burst and only at certain rather early time instants. 
On a diagram (fig.\,\ref{fig:all_polarizations}) we plot MASTER GRB\,150301B polarization low limit  in comparison with other known earlier measurements of the GRB polarization.  At the moment, it is the earliest measurement (from among successful) optical linear polarization. The observation is in agreement with other measurements and shows the ability to burst to have a high degree of optical polarization in the earliest time.

We plan to defer the theoretical and phenomenological description of the nature of prompt optical 
emission of gamma-ray bursts and its polarization by several years, after sufficient amount of 
positive polarimetric observations of gamma-ray bursts becomes available. Our optimism is 
inspired by the doubling of the number of MASTER network telescopes deployed in 2015 in such 
sites of excellent astroclimate as Canary islands, South Africa, and Argentina

\section{Acknowledgements.}

MASTER project is supported in part by the program of
Development of Lomonosov Moscow State University. 

We are grateful to the anonymous referee for valuable discussions and useful changes.

This
work was partically supported by the
Russian Foundation of Fundamental Research 15-02-07875 and 14-02-31546 grants.

DB, AYK, SP and MK  acknowledges support from the National Research Foundation (NRF) of South Africa.

This  work  was also partically supported by state order of  Ministry of Education and
Science of the Russian Federation No. 3.615.2014/K.

We acknowlege excellent support from the CAHA staff, F.
Hoyos and J. Aceituno (CAHA) and the support of the Spanish Ministry
Project AYA 2012-39727.

%
%
%

\bibliographystyle{mn2e}

\bibliography{two_grb}

\begin{thebibliography}{25}
\expandafter\ifx\csname natexlab\endcsname\relax\def\natexlab#1{#1}\fi

\bibitem[{{Buckley} {et~al}\mbox{.}(2015){Buckley}, {Potter}, {Kniazev},
  {Kotze}, {Gorbovskoy}, {Lipunov}, {Kornilov}, {Tyurina}, {Balanutsa},
  {Kuznetsov}, {Chazov}, {Kuvshinov}, {Yurkov}, {Sergienko}, {Varda},
  {Sinyakov}, {Tlatov}, {Parhomenko}, {Dormidontov}, {Sennik}, {Krushinsky},
  {Zalozhnih}, {Popov}, {Levato}, {Saffe}, {Mallamaci}, {Lopez}, \&
  {Podest}}]{gcn17518}
{Buckley} D. {et~al.}, 2015, GRB Coordinates Network, 17518, 1

\bibitem[{{Covino} {et~al}\mbox{.}(2004){Covino}, {Ghisellini}, {Lazzati}, \&
  {Malesani}}]{Covino2004}
{Covino} S., {Ghisellini} G., {Lazzati} D., {Malesani} D., 2004, in
  Astronomical Society of the Pacific Conference Series, Vol. 312, Gamma-Ray
  Bursts in the Afterglow Era, {Feroci} M., {Frontera} F., {Masetti} N., {Piro}
  L., eds., p. 169

\bibitem[{{de Ugarte Postigo} {et~al}\mbox{.}(2015){de Ugarte Postigo},
  {Kruehler}, {Flores}, \& {Fynbo}}]{gcn17523}
{de Ugarte Postigo} A., {Kruehler} T., {Flores} H., {Fynbo} J.~P.~U., 2015, GRB
  Coordinates Network, 17523, 1

\bibitem[{{de Ugarte Postigo} \& {Tomasella}(2015)}]{gcn17710}
{de Ugarte Postigo} A., {Tomasella} L., 2015, GRB Coordinates Network, 17710, 1

\bibitem[{{Gorbovskoy} {et~al}\mbox{.}(2013){Gorbovskoy}, {Lipunov},
  {Kornilov}, {Belinski}, {Kuvshinov}, {Tyurina}, {Sankovich}, {Krylov},
  {Shatskiy}, {Balanutsa}, {Chazov}, {Kuznetsov}, {Zimnukhov}, {Shumkov},
  {Shurpakov}, {Senik}, {Gareeva}, {Pruzhinskaya}, {Tlatov}, {Parkhomenko},
  {Dormidontov}, {Krushinsky}, {Punanova}, {Zalozhnyh}, {Popov}, {Burdanov},
  {Yazev}, {Budnev}, {Ivanov}, {Konstantinov}, {Gress}, {Chuvalaev}, {Yurkov},
  {Sergienko}, {Kudelina}, {Sinyakov}, {Karachentsev}, {Moiseev}, \&
  {Fatkhullin}}]{Gorbovskoy2013}
{Gorbovskoy} E.~S. {et~al.}, 2013, Astronomy Reports, 57, 233

\bibitem[{{Gorbovskoy} {et~al}\mbox{.}(2012){Gorbovskoy}, {Lipunova},
  {Lipunov}, {Kornilov}, {Belinski}, {Shatskiy}, {Tyurina}, {Kuvshinov},
  {Balanutsa}, {Chazov}, {Kuznetsov}, {Zimnukhov}, {Kornilov}, {Sankovich},
  {Krylov}, {Ivanov}, {Chvalaev}, {Poleschuk}, {Konstantinov}, {Gress},
  {Yazev}, {Budnev}, {Krushinski}, {Zalozhnich}, {Popov}, {Tlatov},
  {Parhomenko}, {Dormidontov}, {Senik}, {Yurkov}, {Sergienko}, {Varda},
  {Kudelina}, {Castro-Tirado}, {Gorosabel}, {S{\'a}nchez-Ram{\'{\i}}rez},
  {Jelinek}, \& {Tello}}]{Gorbovskoy2012}
---, 2012, \mnras, 421, 1874

\bibitem[{{Granot}(2003)}]{Granot2003}
{Granot} J., 2003, \apjl, 596, L17

\bibitem[{{Ivanov} {et~al}\mbox{.}(2015){Ivanov}, {Yazev}, {Budnev}, {Gres},
  {Chuvalaev}, {Poleshchuk}, {Gorbovskoy}, {Lipunov}, {Tyurina}, {Kornilov},
  {Balanutsa}, {Kuznetsov}, {Kuvshinov}, {Buckley}, {Potter}, {Kniazev},
  {Kotze}, {Tlatov}, {Parhomenko}, {Dormidontov}, {Sennik}, {Yurkov},
  {Sergienko}, {Varda}, {Sinyakov}, {Krushinski}, {Zalozhnih}, {Popov},
  {Levato}, {Saffe}, {Mallamaci}, {Lopez}, \& {Podest}}]{gcn17689}
{Ivanov} K. {et~al.}, 2015, GRB Coordinates Network, 17689, 1

\bibitem[{{Kornilov} {et~al}\mbox{.}(2012){Kornilov}, {Lipunov}, {Gorbovskoy},
  {Belinski}, {Kuvshinov}, {Tyurina}, {Shatsky}, {Sankovich}, {Krylov},
  {Balanutsa}, {Chazov}, {Kuznetsov}, {Zimnuhov}, {Senik}, {Tlatov},
  {Parkhomenko}, {Dormidontov}, {Krushinsky}, {Zalozhnyh}, {Popov}, {Yazev},
  {Budnev}, {Ivanov}, {Konstantinov}, {Gress}, {Chvalaev}, {Yurkov},
  {Sergienko}, \& {Kudelina}}]{Kornilov2012}
{Kornilov} V.~G. {et~al.}, 2012, Experimental Astronomy, 33, 173

\bibitem[{{Lien} {et~al}\mbox{.}(2015){Lien}, {Burrows}, {Chester}, {Cummings},
  {de Pasquale}, {Maselli}, {Page}, {Palmer}, \& {Siegel}}]{gcn17515}
{Lien} A.~Y. {et~al.}, 2015, GRB Coordinates Network, 17515, 1

\bibitem[{{Lipunov} {et~al}\mbox{.}(2010){Lipunov}, {Kornilov}, {Gorbovskoy},
  {Shatskij}, {Kuvshinov}, {Tyurina}, {Belinski}, {Krylov}, \& {et
  al}}]{lipunov_etal2010}
{Lipunov} V. {et~al.}, 2010, Advances in Astronomy, article id. 349171

\bibitem[{{Lipunov} {et~al}\mbox{.}(2007){Lipunov}, {Kornilov}, {Krylov},
  {Tyurina}, {Belinskii}, {Gorbovskoi}, {Kuvshinov}, {Gritsyk}, {Antipov},
  {Borisov}, {Sankovich}, {Vladimirov}, {Vybornov}, \&
  {Kuznetsov}}]{Lipunov2007ARep}
{Lipunov} V.~M. {et~al.}, 2007, Astronomy Reports, 51, 1004

\bibitem[{{Lipunov} {et~al}\mbox{.}(2004){Lipunov}, {Krylov}, {Kornilov},
  {Borisov}, {Kuvshinov}, {Belinsky}, {Kuznetsov}, {Potanin}, {Antipov},
  {Tyurina}, {Gorbovskoy}, \& {Chilingaryan}}]{Lipunov2004AN}
---, 2004, Astronomische Nachrichten, 325, 580

\bibitem[{{Markwardt} {et~al}\mbox{.}(2015){Markwardt}, {Chester}, {Kennea},
  {Krimm}, {Pagani}, {Page}, {Palmer}, \& {Ukwatta}}]{gcn17688}
{Markwardt} C.~B., {Chester} M.~M., {Kennea} J.~A., {Krimm} H.~A., {Pagani} C.,
  {Page} K.~L., {Palmer} D.~M., {Ukwatta} T.~N., 2015, GRB Coordinates Network,
  17688, 1

\bibitem[{{Maselli} {et~al}\mbox{.}(2015){Maselli}, {Sbarufatti}, {Burrows},
  {Kennea}, {Amaral-Rogers}, {Osborne}, {Page}, {Melandri}, {D'Avanzo}, \&
  {Lien}}]{gcn17533}
{Maselli} A. {et~al.}, 2015, GRB Coordinates Network, 17533, 1

\bibitem[{{M{\'e}sz{\'a}ros} \& {Rees}(1999)}]{meszaros_rees1999}
{M{\'e}sz{\'a}ros} P., {Rees} M.~J., 1999, \mnras, 306, L39

\bibitem[{{Pruzhinskaya} {et~al}\mbox{.}(2014){Pruzhinskaya}, {Krushinsky},
  {Lipunova}, {Gorbovskoy}, {Balanutsa}, {Kuznetsov}, {Denisenko}, {Kornilov},
  {Tyurina}, {Lipunov}, {Tlatov}, {Parkhomenko}, {Budnev}, {Yazev}, {Ivanov},
  {Gress}, {Yurkov}, {Gabovich}, {Sergienko}, \& {Sinyakov}}]{Pruzhinskaya2014}
{Pruzhinskaya} M.~V. {et~al.}, 2014, \na, 29, 65

\bibitem[{{Sanchez-Ramirez} {et~al}\mbox{.}(2015){Sanchez-Ramirez},
  {Castro-Tirado}, {Hoyos}, \& {Aceituno}}]{gcn17697}
{Sanchez-Ramirez} R., {Castro-Tirado} A.~J., {Hoyos} F., {Aceituno} J., 2015,
  GRB Coordinates Network, 17697, 1

\bibitem[{{Steele} {et~al}\mbox{.}(2009){Steele}, {Mundell}, {Smith},
  {Kobayashi}, \& {Guidorzi}}]{Steele2009}
{Steele} I.~A., {Mundell} C.~G., {Smith} R.~J., {Kobayashi} S., {Guidorzi} C.,
  2009, \nat, 462, 767

\bibitem[{{Tyurina} {et~al}\mbox{.}(2015){Tyurina}, {Gorbovskoy}, {Kornilov},
  {Balanutsa}, {Kuznetsov}, {Kuvshinov}, {Ivanov}, {Yazev}, {Budnev}, {Gres},
  {Chuvalaev}, {Poleshchuk}, {Buckley}, {Potter}, {Kniazev}, {Kotze}, {Tlatov},
  {Parhomenko}, {Dormidontov}, {Sennik}, {Yurkov}, {Sergienko}, {Varda},
  {Sinyakov}, {Krushinski}, {Zalozhnih}, {Popov}, {Levato}, {Saffe},
  {Mallamaci}, {Lopez}, \& {Podest}}]{gcn17690}
{Tyurina} N. {et~al.}, 2015, GRB Coordinates Network, 17690, 1

\bibitem[{{Uehara} {et~al}\mbox{.}(2012){Uehara}, {Toma}, {Kawabata},
  {Chiyonobu}, {Fukazawa}, {Ikejiri}, {Inoue}, {Itoh}, {Komatsu}, {Miyamoto},
  {Mizuno}, {Nagae}, {Nakaya}, {Ohsugi}, {Sakimoto}, {Sasada}, {Tanaka},
  {Uemura}, {Yamanaka}, {Yamashita}, {Yamazaki}, \& {Yoshida}}]{Uehara2012}
{Uehara} T. {et~al.}, 2012, \apjl, 752, L6

\bibitem[{{Ukwatta} {et~al}\mbox{.}(2015){Ukwatta}, {Barthelmy}, {Baumgartner},
  {Cummings}, {Gehrels}, {Krimm}, {Lien}, {Markwardt}, {Palmer}, {Sakamoto},
  {Stamatikos}, \& {Tueller}}]{gcn17524}
{Ukwatta} T.~N. {et~al.}, 2015, GRB Coordinates Network, 17524, 1

\bibitem[{{Vestrand} {et~al}\mbox{.}(2005){Vestrand}, {Wozniak}, {Wren},
  {Fenimore}, {Sakamoto}, {White}, {Casperson}, {Davis}, {Evans}, {Galassi},
  {McGowan}, {Schier}, {Asa}, {Barthelmy}, {Cummings}, {Gehrels}, {Hullinger},
  {Krimm}, {Markwardt}, {McLean}, {Palmer}, {Parsons}, \&
  {Tueller}}]{Vestrand2005}
{Vestrand} W.~T. {et~al.}, 2005, \nat, 435, 178

\bibitem[{{Willingale} {et~al}\mbox{.}(2013){Willingale}, {Starling},
  {Beardmore}, {Tanvir}, \& {O'Brien}}]{Willingale2013}
{Willingale} R., {Starling} R.~L.~C., {Beardmore} A.~P., {Tanvir} N.~R.,
  {O'Brien} P.~T., 2013, \mnras, 431, 394

\bibitem[{{Yonetoku} {et~al}\mbox{.}(2011){Yonetoku}, {Murakami}, {Gunji},
  {Mihara}, {Toma}, {Sakashita}, {Morihara}, {Takahashi}, {Toukairin},
  {Fujimoto}, {Kodama}, {Kubo}, \& {IKAROS Demonstration Team}}]{Yonetoku2011}
{Yonetoku} D. {et~al.}, 2011, \apjl, 743, L30

\end{thebibliography}

\newpage

\begin{figure}
\includegraphics[width=84mm]{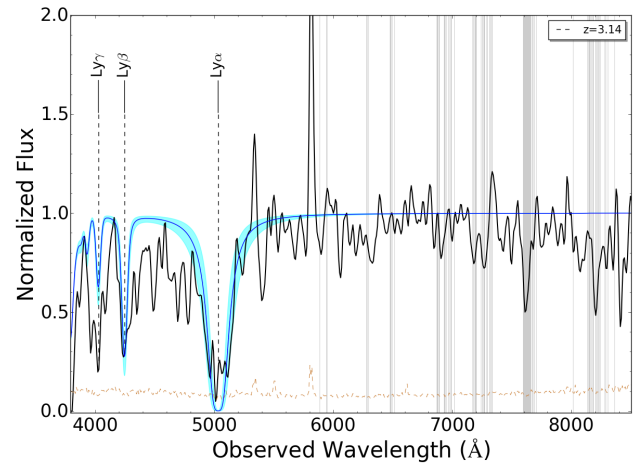}
\caption{
GRB\,150413A spectrum obtained by GTC.  $z = 3.14$.
On the plot, vertical gray lines mark telluric absorption, black spectrum correspond to the signal and the brown one is the $1 \sigma$ error. Over-plotted blue line and cyan area is a Lyman series line fitting with $log (N_H) =  22.0 \pm 0.25$
 \label{fig:grb150413A_spectrum}}
\end{figure}

\begin{table}
 \centering \caption{Photometry data for GRB\,150301B by MASTER in the
unfiltered band with two orthogonal polarizing filters.}

\label{tab:polarization150301B}
  \begin{tabular}{@{}cccc@{}}
\hline
\hline
$t-T_{trigger}$&Exp.&Unfiltered $P_{45}$&Unfiltered $P_{135}$ \\
 (s)  &       (s)     & (mag)      &(mag) \\
\hline
79 & 10 & $ 15.26  \pm  0.04 $ & $15.43  \pm  0.05$\\
127 & 20 & $ 15.61  \pm  0.05$ & $15.71  \pm  0.05$\\
187 & 40 & $16.17  \pm  0.06 $& $16.18  \pm  0.06$\\
276 & 50 & $16.47  \pm  0.08 $& $16.54  \pm  0.07$\\
389 & 80 & $17.14  \pm  0.10 $& $16.79  \pm  0.08$\\
502 & 100 &$ 17.48  \pm  0.12$ & $17.49  \pm  0.10$\\
659 & 130 &$ 17.89  \pm  0.14$ & $17.71  \pm  0.10$\\
827 & 160 &$ 18.13  \pm  0.16$ & $18.22  \pm  0.12$\\
1015 & 180 &$ 18.38  \pm  0.17$ &$ 18.16  \pm  0.12$\\
1246 & 180 &$ 18.48  \pm  0.18$ &$ 18.86  \pm  0.14$\\
1465 & 180 &$ 19.01  \pm  0.21 $&$ 18.50  \pm  0.13$\\
1688 & 180 &$ 18.68  \pm  0.19 $&$ 18.85  \pm  0.14$\\
1914 & 180 &$ 18.98  \pm  0.21 $&$ 18.87  \pm  0.13$\\
2142 & 180 &$ 19.03  \pm  0.19 $&$ 19.09  \pm  0.14$\\

\hline     
\end{tabular}
\end{table}

\begin{table}
 \centering \caption{Photometry data for GRB\,150413A by MASTER in the
unfiltered band with two orthogonal polarizing filters.}

\label{tab:polarization150413A}
  \begin{tabular}{@{}cccccc@{}}
\hline
\hline
$t-T_{trigger}$&Exp.& Filter 1 &Magnitude& Filter 2 & Magnitude \\
 (s)  &       (s)  &   & (mag)   &   &(mag) \\
\hline
131 & 30 & $P|$ & $17.88 \pm 0.16$  & P- & $17.92 \pm 0.12$ \\
223 & 40 & $P|$ & $16.93 \pm 0.09$  & P- & $17.31 \pm 0.10$ \\
323 & 60 & $P|$ & $16.29 \pm 0.05$  & P- & $16.18 \pm 0.07$ \\
435 & 90 & $P|$ & $15.66 \pm 0.03$  & P- & $15.63 \pm 0.05$ \\
827 & 170 & $P|$ & $14.91 \pm 0.01$  & P- & $14.92 \pm 0.04$ \\
1062 & 180 & $P|$ & $14.92 \pm 0.01$  & P- & $14.97 \pm 0.04$ \\
1299 & 180 & $P|$ & $15.08 \pm 0.01$  & P- & $15.09 \pm 0.04$ \\
1546 & 180 & $P|$ & $15.23 \pm 0.02$  & P- & $15.25 \pm 0.05$ \\
1784 & 180 & $P|$ & $15.38 \pm 0.02$  & P- & $15.36 \pm 0.05$ \\
2023 & 180 & $P|$ & $15.48 \pm 0.02$  & P- & $15.50 \pm 0.05$ \\
2265 & 180 & $P|$ & $15.57 \pm 0.02$  & P- & $15.57 \pm 0.05$ \\
2509 & 180 & $P|$ & $15.70 \pm 0.03$  & P- & $15.73 \pm 0.06$ \\
2751 & 180 & $P|$ & $15.75 \pm 0.03$  & P- & $15.75 \pm 0.06$ \\
2997 & 180 & $P|$ & $15.90 \pm 0.03$  & P- & $15.80 \pm 0.06$ \\
3240 & 180 & $P|$ & $16.00 \pm 0.04$  & P- & $15.94 \pm 0.06$ \\
3475 & 180 & $P|$ & $16.16 \pm 0.04$  & P- & $16.03 \pm 0.06$ \\
3713 & 180 & $P|$ & $16.13 \pm 0.04$  & P- & $16.08 \pm 0.07$ \\
3943 & 180 & $P|$ & $16.23 \pm 0.05$  & P- & $16.22 \pm 0.07$ \\
4183 & 180 & $P|$ & $16.29 \pm 0.05$  & P- & $16.27 \pm 0.07$ \\
4426 & 180 & $P|$ & $16.31 \pm 0.05$  & P- & $16.41 \pm 0.07$ \\
4673 & 180 & $P|$ & $16.35 \pm 0.05$  & P- & $16.31 \pm 0.07$ \\
4913 & 180 & $P|$ & $16.41 \pm 0.06$  & P- & $16.49 \pm 0.08$ \\
5152 & 180 & $P|$ & $16.46 \pm 0.06$  & P- & $16.42 \pm 0.07$ \\
5393 & 180 & $P|$ & $16.42 \pm 0.06$  & P- & $16.69 \pm 0.08$ \\
5627 & 180 & $P|$ & $16.49 \pm 0.06$  & P- & $16.39 \pm 0.07$ \\
5871 & 180 & $P|$ & $16.62 \pm 0.07$  & P- & $16.42 \pm 0.07$ \\
6120 & 180 & $P|$ & $16.62 \pm 0.07$  & P- & $16.55 \pm 0.08$ \\
6363 & 180 & $P|$ & $16.67 \pm 0.07$  & P- & $16.76 \pm 0.08$ \\
6603 & 180 & $P|$ & $16.72 \pm 0.07$  & P- & $16.64 \pm 0.08$ \\

\hline     
\end{tabular}
\end{table}

\end{document}